\documentclass[aps,prd,twocolumn,letterpaper,floatfix,showpacs,amsmath]{revtex4} 
\newlength{\picwidth}
 
\usepackage[dvips]{graphicx}

\newcommand{\ern}{\mathcal{E}}
\begin{document}

\renewcommand{\thefigure}{\arabic{figure}}
\title{
Spacetime Encodings III - Second Order Killing Tensors.}
\author{Jeandrew Brink}
\affiliation{Theoretical Astrophysics, California Institute of Technology, Pasadena, CA 91103  }

\begin{abstract}
This paper explores the Petrov type D, stationary axisymmetric vacuum (SAV) spacetimes that were found by Carter to have separable Hamilton-Jacobi equations, and thus admit a second-order Killing tensor.
The derivation of the spacetimes presented in this paper borrows from ideas about dynamical systems, and illustrates concepts that can be generalized to higher- order Killing tensors. The relationship between the components of the Killing equations and metric functions are given explicitly. The origin of the four separable coordinate systems found by Carter is explained and classified in terms of the analytic structure associated with the Killing equations. A geometric picture of what the orbital invariants may represent is built.  Requiring that a SAV spacetime admits a second-order Killing tensor is very restrictive, selecting very few candidates from the group of all possible SAV spacetimes. This restriction arises due to the fact that the  consistency conditions associated with the Killing equations
require that the field variables obey a second-order differential equation, as opposed to a fourth-order differential equation that imposes the weaker condition that the spacetime be SAV.  This paper introduces ideas that could lead to the explicit computation of more general orbital invariants in the form of higher-order Killing Tensors.

\end{abstract}
\pacs{ }

\maketitle

\section{Introduction}
It is well known that many Petrov Type D spacetimes admit second order Killing tensors \cite{CarterSeparability,Walker,MathTheoryofBlackHoles} 
 . These  Killing tensors give rise to constants of geodesic motion that can be used to describe particle motion within the spacetimes and so provide clues to the spacetime's  experimental or observational signature  \cite{JdB0, WaveFormMachineSteveDrasco}.

 This paper explores the existence of second-order Killing tensors in the restricted context of stationary axisymmetric vacuum (SAV)  spacetimes. The aim is to generate the ideas necessary to help gain explicit control over the geodesic structure of all  SAV spacetimes, if this is possible.

Much of the more recent work done on second-order Killing tensors is written in terms of the repeated principal null directions associated with the Weyl tensor of Type D spacetimes.  Once the results are postulated within this formalism it is easy to show that they are correct as is elegantly done in \cite{Walker,MathTheoryofBlackHoles}. It is however unclear how to generalize this work to higher order Killing tensors, where the solutions are not a priori known.
A constructive derivation for finding such Killing tensors, similar to Carter's original 30 page derivation \cite{CarterSeparability} for the second-order case, is thus required.
With the advantage of hindsight it should be noted that the second order Killing tensor problem can effectively be reduced to a linear problem \cite{JdB1}. This explains the early success of Carter's direct approach to finding solutions. This feature does not extend to higher order Killing tensors, where the problem becomes distinctly nonlinear \cite{JdB1}. Despite this difficulty, a number of the features observed in the derivation of spacetimes admitting second-order Killing tensors do generalize to higher-order Killing tensors.
     
This paper aims to familiarize the reader with a constructive approach to finding second-order Killing tensors, taking special care to highlight the features and results of such a derivation that  can be extended to higher-order Killing tensors. While none of the results presented in this paper are new, the derivations I believe are and the insights gained have been conducive to seeking and understanding solutions to the higher-order Killing tensor equations in SAV spacetimes.  The purpose of this paper is mainly pedagogical. It serves as a prototype calculation that will be fleshed out  in subsequent papers \cite{JdB3, JdBZV1, JdBZV2} to provide a formalism for  checking for the existence of and subsequently writing down a formal solution for the components of fourth-order Killing tensors. 

The restrictions of stationarity and axisymmetry result in two Killing vectors on the general SAV spacetime manifold. If these Killing vectors commute, the vacuum spacetime is entirely determined by a complex Ernst potential \cite{JdB0}. In this context, the problem of finding the existence of a second-order Killing tensor simplifies. As is shown in Sec. \ref{SEC2}, it  essentially  reduces to  several two-degree-of-freedom problems \cite{JdB1}. An investigator can  benefit from the work performed in this field, and the related field of super-integrable systems  \cite{Koenigs,Whit,WolfT,Hall,Hietarinta,Kalnins} to build up intuition about how the field equations and Killing equations interrelate. Two-degree-of-freedom dynamical systems have the great advantage that they can be easily visualized \cite{JdB1}; furthermore, Killing tensors take on a distinct geometric meaning that was discussed in \cite{JdB1} and this can be used to  aid calculation. It is this structure, studied in this and subsequent work  \cite{JdB3,JdBZV1}  that admits generalization to higher-order Killing tensors. 

This derivation relies heavily on the techniques developed in the field of dynamical systems and direct searches for invariants \cite{Hall,Hietarinta}. It should however not be forgotten  that what one is actually describing is the physical trajectory of an observer through four dimensional spacetime. This observer has his/her own coordinate system and can perform experiments to determine the nature of the spacetime through which he/she is traveling. As will be shown in Sec. \ref{SEC2}, the existence of second-order Killing tensors implies not only separability of the metric functions, but the existence of special coordinate systems, in which separability occurs (see Sec \ref{SECSEP}) . How these arise and their implications will be discussed in Secs.~\ref{SEC2} and~\ref{SECCON}.

The approach taken in the constructive derivation given here is direct; simply solve both the field and Killing equations simultaneously and see how much you could possibly learn. For simplicity, the approach is coordinate-based, with an emphasis on identifying the structures that relate to work done in two-degree-of-freedom dynamical systems.
A formulation on an  appropriately chosen tetrad basis is given in \cite{JdB3} which  simplifies the analysis somewhat for higher-order Killing tensors. The tetrad formalism however disguises other properties of note. A particular advantage of the formulation given here is that it is immediately apparent how to write down the explicit components of the Killing tensor given the metric in the separable coordinate system (Sec \ref{SEC2}).  

For the sake of completeness, the relationship between general type D metrics and those admitting second-order Killing tensors is given in Sec \ref{SECDDD}. Finally this paper is concluded by highlighting which features of this particular derivation generalize to higher-order Killing tensors.

\section{Second Order Killing Tensors on SAV spacetimes}
\label{SEC2}
Consider a totally symmetric tensor $T^{(\alpha_1\alpha_2)}$ of order $2$ on a SAV spacetime with line element,
\begin{align}
ds^2 &=  e^{-2\psi}\left[e^{2\gamma}(d\rho^2+dz^2)+R^2d\phi^2\right]-e^{2\psi}(dt-\omega d\phi)^2 \label{LineEle}
\end{align}
where the functions $\psi$, $\gamma$, $\omega$ and $R$ are determined by the complex Ernst potential $\ern$ \cite{JdB1}.

For purposes of calculation, the components will be split into two parts. Allow the indices $A,B$ to run over $\{t,\phi\}$ and the indices $i,j$ over  $\{\rho,z \}$, while the indices $\alpha_i$ run over all four possibilities.

If a tensor  $T$ obeys the Killing equation,
\begin{align}
T^{(\alpha_1\alpha_2;\alpha_{3})}=0, \label{KillEQ22}
\end{align}
it can be shown that the quantity
\begin{align}
T^{(\alpha_1\alpha_2)}p_{\alpha_1}  p_{\alpha_2}= Q
\end{align}
is constant along a geodesic of a particle with four momentum $p_\alpha$ and so provides a constant of motion. A trivial solution to \eqref{KillEQ22} is found by replacing $T$ with the metric $g$; then the constant $Q$ is the rest mass~$-\mu ^2$. To fully describe the geodesic in four dimensions, we need three more distinct quantities that remain constant along the curve to use as coordinates ideally suited to the curve. If we assume these quantities are second order Killing tensors,  two reducible Killing tensors can be constructed from the Killing vectors $\partial_t$ and $\partial_\phi$. We are searching for the third and a method of computing its explicit form. 

Start by writing out the Killing equations.  Let \mbox{$V=e^{2\gamma-2\psi}$}; then the non-zero Christoffel symbols associated with the metric \eqref{LineEle} are 
\begin{align}
\Gamma^\rho_{\rho\rho}&= -\Gamma^\rho_{zz}=\Gamma^z_{z\rho}= 2 \partial_\rho (\ln V), & \Gamma^A_{Bi}&= \frac{1}{2}g^{AD} \partial_i g_{BD},\notag\\
\Gamma^z_{zz}&= -\Gamma^z_{\rho\rho}=\Gamma^\rho_{z\rho}= 2 \partial_z ( \ln V),&\Gamma^i_{BC}&=-\frac{1}{2V}\partial_i g_{BC}.
\end{align} 
The absence of explicit dependence on  $t$ and  $\phi$  greatly reduces the number of equations to be considered. The Killing equations that are not trivially satisfied are 
$T^{(AB;j)}=0$ and   $T^{(ij;k)}=0$. The first group of equations,  $T^{(AB;j)}=0$,  define the six gradients of the three $T^{(AB)}$ components: 
\begin{align}
T^{(AB)}_{,j}&=   V\partial_k g^{AB} T^{k j}; \label{KT1}
\end{align}
the second group of equations, $T^{(ij;k)}=0$, decouple and describe the Killing equations for a two dimensional manifold with metric $g_{ij}= V \delta_{ij}$:   
\begin{align}
0&=VT^{\rho\rho}_{,\rho}+V_{,\rho}T^{\rho\rho}+V_{,z}T^{\rho z},\label{KT2a}\\
0&=VT^{\rho \rho}_{,z}+2VT^{\rho z}_{,\rho} +V_{,\rho} T^{\rho z} +V_{,z}T^{zz},\label{KT2b}\\
0&=VT^{zz}_{,\rho}+2VT^{\rho z}_{,z} +V_{,\rho} T^{\rho \rho} +V_{,z}T^{\rho z},\label{KT2c}\\
0&=VT^{zz}_{,z}+V_{,\rho}T^{\rho z}+V_{,z}T^{zz}.\label{KT2d}
\end{align}
This group of equations is entirely equivalent to the geometric picture given in \cite{JdB1}.  The two-dimensional Killing equations represent the Fourier-series expansion of the invariant distinct from the Hamiltonian in phase space. 
 Associated with these equations we find the analytic structure observed by   \cite{JdB1,Hall,Xanth3}. This structure represents a coordinate freedom that can either be viewed as a friend or a foe.  The coordinate freedom in two-degree-of-freedom Hamiltonian systems makes it very difficult to identify whether the system is integrable. In the case of the SAV spacetimes and in this derivation, however, the coordinate freedom is a friend and allows a gauge to be chosen such that the components of the Killing equations can be explicitly written down. To make this gauge freedom explicit, consider the linear combinations of equations \eqref{KT2a}-\eqref{KT2c} and  \eqref{KT2b}-\eqref{KT2d} from which the Cauchy-Riemann conditions  become apparent,
\begin{align}
(T^{\rho\rho}-T^{zz})_{,\rho}&=2T^{\rho z}_{,z},& (T^{\rho\rho}-T^{zz})_{,z}&=-2T^{\rho z}_{,\rho}\label{CRS}
\end{align}
(It is easier to identify the analytic structure if the equations are written in terms of a tetrad, as will be done in~\cite{JdB3}.)

It is natural to define the complex variable \mbox{$t=T^{\rho\rho}-T^{zz}+2iT^{\rho z}$} and the real variable \mbox{$s = T^{\rho\rho} +T^{zz}$}. In addition, the introduction of \mbox{$\zeta = 1/2(\rho + i z)$} and its complex conjugate $\overline{\zeta}$ as independent variables results in the following expressions for the Killing equations:
\begin{align}
T^{AB}_{,{\zeta}}&=\frac{1}{2}V\left(\overline{t}\partial_{\overline{\zeta}} g^{AB}+s\partial_{{\zeta}}g^{AB}   \right),\label{KillT2TAB}\\
 (Vs)_{,\zeta} &= -\left(\frac{1}{2} V\overline{t}_{,\overline{\zeta}}  +\overline{t}V_{,\overline{\zeta}}\right),\label{KillT2s}\\ 
 t_{,\overline{\zeta}}&=0.\label{KillT2sss}
\end{align}
The complex conjugates of these equations must also hold. Bear in mind that $t(\zeta)$ is an analytic function that admits a power series expansion in $\zeta$. Formally,
\begin{align}
t(\zeta) = \sum_n a_n(\zeta-\zeta_0)^n. \label{PowerT}
\end{align}
The solution method now involves eliminating the variables $T^{AB}$ and $s$ in favor of a higher order equation for $t(\zeta)$ by writing down their integrability conditions. These equations limit the freedom of choice of the coefficients $a_n$, which determine the analytic function $t$.

For the Eqs. \eqref{KillT2TAB} and \eqref{KillT2s}, after some algebra we are respectively left with
\begin{align}
\left( \frac{\left((V g^{AB})^2\overline{t}\right)_{,\overline{\zeta}} }{ V g^{AB} }  \right)_{,\overline{\zeta}}=CC, \label{KillT2TABint}
\end{align}
and
\begin{align}
 \left( \frac{\left(V^2 \overline{t}\right)_{,\overline{\zeta}}}{V}\right)_{,\overline{\zeta}}=CC, \label{kill2Standard}
\end{align}
where the notation ``$=CC$'' should be read as ``equals its complex conjugate''. There are four different equations of the same form,
indicating the hunt for four different two-manifolds admitting a second-order Killing tensor.  

Only very special potentials $V$ admit a solution. A general method for solving Eq. \eqref{kill2Standard} was suggested by Hall \cite{Hall} and proceeds as follows:  Make the conformal transformation generated by the analytic function \mbox{$r=r_1+ir_2$}, namely that $d\zeta = rd\tilde{\zeta}$ with $\tilde{\zeta}=1/2(\tilde{\rho}+i\tilde{z})$.  Furthermore, choose $r$ such that $t = 1/2 r^2$.
Then in the new coordinate system Eq. \eqref{kill2Standard} becomes 
\begin{align}
\frac{1}{\overline{r}}\partial_{\overline{\tilde{\zeta}\tilde{\zeta}
}}(\overline{r}V)=CC.  
\end{align}
Multiplying through by $r\overline{r}$ and using the fact that $r_{,\overline{\tilde{\zeta}}}=0$, we obtain
\begin{align}
\partial_{\tilde{\rho}\tilde{z}} (r\overline{r}V)=0. \label{EQKILLSEP}
\end{align}
The general solution can be written down as
\begin{align}
r\overline{r} V = f_1(\tilde{\rho})+f_2(\tilde{z}),\label{KILLT4VVVB}
\end{align}
where the functions $f$ are arbitrary and must be chosen such that the field equations are also satisfied.

Eq. \eqref{KillT2TABint}  has the same form, and  the general solution for $g^{AB}$ can also be written down in terms of another set of free functions $f^{AB}$, 
\begin{align}
 { r\overline{r}}V g^{AB} = f_1^{AB}(\tilde{\rho})+f_2^{AB}(\tilde{z}).
 \label{KILLT4VV}
\end{align}
This implies that in the separable coordinate system, the conformally rescaled metric $\hat{g}^{\alpha \beta} =r \overline{r} V g^{\alpha \beta}$ is separable in all components. As a result, the Hamilton-Jacobi equation associated with this Hamiltonian is also separable.

In the new coordinate system, the Killing Eqs.~\eqref{KillT2TAB} and~\eqref{KillT2s} take on a greatly simplified form:
\begin{align}
\partial_{{\tilde{\zeta}}}\left(T^{AB}    -\frac{1}{2} V s g^{AB}\right)  &=+\frac{1}{4}  \partial_{\overline{\tilde{\zeta}}}\left(r\overline{r} V    g^{AB}\right),\notag   \\
 (Vs)_{,\tilde{\zeta}} &= - \frac{1}{2} \partial_{\overline{\tilde{\zeta}}}\left(r\overline{r} V \right). \label{KEEEQ}
\end{align}
The separable form of the quantities on the right allow the solution to be written down explicitly,
\begin{align}
V s &= -\frac{1}{2}( f_1(\tilde{\rho})-f_2(\tilde{z}))+a_s,\notag\\
T^{AB} &= \frac{1}{2} V s g^{AB} + \frac{1}{4} ( f_1^{AB}(\tilde{\rho})-f_2^{AB}(\tilde{z})) +a_t^{AB}, \label{KILLT4s}
\end{align}
where $a_t$ and $a_s$ are real constants.

It should be observed that the potential $V$ and the metric components must obey a second-order differential equation in some coordinate system, if they are also to admit  a second-order Killing tensor. The $V's$ generated for SAV spacetimes are in fact only required to obey  fourth order non-linear differential equations, one form of which is Ernst's equation. The condition that there be a second-order Killing tensor on the spacetime is therefore far more restrictive than the conditions governing the generation of the spacetimes themselves. It is shown in \cite{JdB3} that this condition limits the Petrov Type to D.

This example also illustrates the difficulty in identifying whether a particular Hamiltonian is integrable \cite{JdB1,Hietarinta}. Although,  as shown here, the potential $V$ has a very succinct form in some coordinate system, the coordinate transformation to this system is in general not known. This greatly hinders the identification of whether a sample Hamiltonian admits a second order Killing tensor.  

In the case of the SAV field equations this particular difficulty can be overcome. The analytic function $t$ is related to the choice of gauge function $R(\rho , z)$ of the metric~\eqref{LineEle}.
To see this, expand  $(\eqref{KillT2TABint}- g^{AB}\eqref{kill2Standard})/V$ to yield the three equations involving the metric components $g^{AB}$,
\begin{align} 
3 \overline{t}_{,\overline{\zeta}} \partial_{\overline{\zeta}}g^{AB} + 2\overline{t}(\partial_{\overline{\zeta\zeta}}g^{AB}+2(\ln V)_{,\overline{\zeta}}\partial_{\overline{\zeta}}g^{AB}) = CC; \label{KILL3}
\end{align}
now take the ``trace'' of this equation by multiplying by $g_{AB}$ and summing over $A$ and $B$. Recall that \mbox{$g^{AB}g_{AB}=2$} and  $\det(g_{AB})=-R^2$, where $R_{,\zeta\overline{\zeta} } =0$. Furthermore using the SAV field equations (App.~\ref{FieldEQ}) it can be shown that the ``trace'' of
 \eqref{KILL3} becomes;
\begin{align} 
M_2 \overline{t}_{,\overline{\zeta}}  +2 \overline{t}M_{2,\overline{\zeta}}  = CC, \label{TTRANS}
\end{align}
where $M_2= (\ln R)_{,\overline{\zeta}}$. If we choose the coordinate system to be $R(\rho,z)=\rho$ and correspondingly $M_2=1/\rho$, $M_{2,\overline{\zeta}} = -1/\rho^2$,  then Eq.~\eqref{TTRANS} reduces to 
\begin{align}
\overline{t}_{,\overline{\zeta}} - 2\overline{t}/\rho=CC; \label{TTRANS2} \end{align}  
repeated differentiation yields $\overline{t}_{,\overline{\zeta\zeta\zeta}}=0$ and as a result
\begin{align}
{t}=a_2 {\zeta}^2+ia_1{\zeta}+a_0, \label{TTTEQ}
\end{align} where the $a$'s are real constants.

Any rescaling of the coordinates and $a$'s can be absorbed in the as yet undetermined metric functions. So one needs only consider the different types of transformations that result from the different natures of the roots of $t$. There are four possibilities:
\begin{itemize}
\item[]{A) $t$ is constant,  $a_0 =\frac{1}{2}$, $a_1=a_2=0$;}
\item[]{B) $t$ is linear,  $a_0 = 0$, $a_1 = 1$, $a_2=0$ ;}
\item[]{C) $t$ is quadratic with a double root, $a_1= a_0 =0$, $a_2 =\frac{1}{2}$  ;}
\item[]{D) $t$ is quadratic with two distinct roots, $a_0 \neq 0$,  \mbox{$a_1=0$}, $a_2 =2$.}
\end{itemize}
Given these four possibilities and the  expression for $t$, 
the corresponding transformation to a separable system generated by 
$r = 2 t^{1/2}$ can be found. The field equations can now be solved 
and the explicit functional form obtained for the metric coefficients. 
\begin{table*}[hbt] 
\begin{align}
\begin{array}{|cc|c|c|c|c|c|c|}\hline\hline
 \multicolumn{2 }{|c| }{ \mbox{Analytic Struct.}}&t &\tilde{\zeta} &r&x&y& R 
 \\\hline
(A)&\mbox{Constant }&\frac{1}{2}&\zeta & 1& &&  \tilde{\rho} 
\\ \hline
(B)&\mbox{Linear } &i\zeta &  \sqrt{2}e^{-\pi/4i} \zeta^{1/2}&  -i\tilde{\zeta} &  &&  \frac{1}{2}\tilde{\rho}\tilde{z} \\ \hline
(C)&\mbox{Quadratic}&&&&&&\notag\\
&\mbox{single root}&  \frac{1}{2}\zeta^2&\ln \zeta& e^{\tilde{\zeta}}  & e^{\tilde{\rho}/2}    &   \sin(\tilde{z}/2)   & 2 x \sqrt{1-y^2}   
\\ \hline
(D)&\mbox{Quadratic}&&&&&&\notag\\
&\mbox{double root}&  
2(\zeta^2+a_4^2) & \frac{1}{2}\ln\left(\frac{i\zeta}{a_4}+\sqrt{\left(\frac{i\zeta}{a_4}\right)^2-1}\right)& 2a_4 \sinh \tilde{2\zeta} & \cosh \tilde{\rho}  & \cos \tilde{z}& 2a_4\sqrt{(x^2-1)(1-y^2)}     \notag\\ 
 \hline
\end{array}
\end{align}

\begin{align}
\begin{array}{|c|c|c|l|cl|cl|}\hline\hline
\mbox{Analytic Structure}& t &M_2 & \mbox{Functions} \ \ f_1 &&f_2& &\mbox{Constraints} \\ \hline
(A)&\frac{1}{2} &  \frac{1}{\tilde{\rho}}  &  b_4\tilde{\rho}^4 - b_2\tilde{\rho}^2 + b_0 & &  2b_2 \tilde{z}^2 + b_1 \tilde{z} & b_0 &=\frac{b_2^2}{4 b_4}+\frac{b_1^2}{8 b_2}  \\ \hline
(B)& i\zeta &  \frac{1}{\tilde{\rho}} + \frac{i}{\tilde{z}}    & b_4\tilde{\rho}^4 + b_2\tilde{\rho}^2 + b_0 && c_4 \tilde{z}^4 + b_2 \tilde{z}^2 &b_0&=\frac{b_2^2}{4 b_4}+\frac{b_2^2}{4 c_4} 
  \\ \hline
(C)& \frac{1}{2}\zeta^2& 1/2\left(1- \frac{iy}{  \sqrt{1-y^2}} \right)  & b_4 x^2 + b_2 x  &&c_4 y +2c_2( y^2+1)& b_4&=\frac{2b_2^2c_2}{16 c_2^2-c_4^2}\\ \hline
(D)& 2(\zeta^2+a_4^2) &  \frac{x}{\sqrt{x^2-1}}+i\frac{y}{\sqrt{1-y^2}}  & b_2(x^2+1) + b_1x &&   c_2(y^2+1) + c_1y&b_1^2&= 4 b_2^2-\frac{b_2 c_1^2}{c_2} + 4 b_2 c_2 \notag\\ \hline
\end{array}
\end{align}

\caption{Four different coordinate transformations \label{fourCor} with accompanying metric functions and constraints. The four transformations originate from Eq. \eqref{TTTEQ}.  The metric functions obey the field equations set out in Appendix \ref{FieldEQ}, and the Killing equations given in Appendix \ref{AppendSep}. The full calculation of the metric functions and constraints as well as the remaining metric functions is performed in Appendix \ref{SeperableFunctions}.}
\end{table*}

Two other linear combinations of $\eqref{KILL3}$ allow us to eliminate the metric fields in favor of their derivatives $M_i$ alone. (The origin of these linear combinations is more easily seen from the tetrad formulation considered in \cite{JdB3}.) Thus, the coupling between the Killing and field equations can be expressed as 
\begin{align}
M_2 \overline{t}_{,\overline{\zeta}}  +2 \overline{t}M_{2,\overline{\zeta}}  = CC,  \notag\\
 \frac{1}{4}\overline{t}_{,\overline{\zeta\zeta}}+\frac{3}{2}\overline{t}_{,\overline{\zeta}}M_A+   \overline{t}(2M_A^2+  M_{A,\overline{{\zeta}}})   =CC,\notag\\
 \frac{1}{2}\overline{t}_{,\overline{\zeta\zeta}} + \frac{3}{2}\overline{t}_{,\overline{\zeta}}M_B+   \overline{t}(M_B^2 +M_{B, \overline{{\zeta}}})     = CC, \notag\\
\frac{3}{2}\overline{t}_{,\overline{\zeta}}M_C+ \overline{t}\left( (2M_A+M_B- M_2)M_C + M_{C ,\overline{{\zeta}}}  \right) = CC, \label{KILLCURVE}
\end{align}
where the $M's$ can be defined 
 in terms of the metric functions;
\begin{align}
M_2 &=(\ln R)_{,\overline{\zeta}},&
M_A &= \frac{1}{2}\partial_{\overline{\zeta}}(\ln V g^{\phi\phi}),\notag\\ 
M_B&= (\ln V)_{,\overline{\zeta}},&
M_C&=- \omega_{,\overline{\zeta}} e^{2\psi} \frac{1}{R}.
\end{align}
These equations are valid in any coordinate system. The field equations that govern the derivatives of the $Ms$ can be found in App.~\ref{FieldEQ}. 

\section{Specialization to the separable coordinate system}
\label{SECSEP}
In order to find the actual metric functions it is computationally preferable to specialize the formalism to the separable coordinate system $(\tilde{\zeta},\overline{\tilde{\zeta}})$. A valid solution of the Killing equations yielding a Killing tensor distinct from the metric is given by $t=\frac{1}{2}$   and  the remaining components are defined by Eqs. \eqref{KILLT4VVVB}  and \eqref{KILLT4VV}  with the specialization that $r=1$. Denote the gradients of the field variables in this coordinate system as $\tilde{M}$. The $\tilde{M}s$ are given in terms of the separable functions in  \eqref{SepMetFunc}.   

In addition to the field Eqs. \eqref{FieldEQMABC} and  \eqref{FieldEQMABCb}, the $\tilde{M}s$ must also satisfy Eqs. \eqref{KILLCURVE}  in the special case when $\tilde{t}$ is a real constant. The resulting coupling conditions governing the field variables are expressed in Eq. \eqref{KillABCD}  . 
There are thus essentially two coordinate systems. In the first the metric function $R=\rho$ is known, and the metric is assumed to admit a second order Killing tensor. In the second, the explicit solution for the Killing-tensor components is known in terms of the metric functions via Eqs. \eqref{KILLT4s}.
The metric functions in this coordinate systems are separable (Eqs. \eqref{KILLT4VVVB}  and \eqref{KILLT4VV}). There are four possible transformations between these two coordinates systems. These transformations are known, can be expressed in terms of the analytic functions $r$, and they with some of the metric functions are tabulated in Table \ref{fourCor}. The computation of the metric functions compatible with both the field and Killing equations is given in App.~\ref{SeperableFunctions}, and the remaining metric functions are listed in Eq.~\eqref{SepMetFuncX}.
  
As a result of the four possible transformations, there are four families of metrics that admit  second-order Killing tensors. The first family (A) includes flat space. The last family (D), which results from the transformation with two distinct zeroes, includes spacetimes such as  Kerr and Schwarzschild  and is thus of astrophysical interest. 

A historical note is that the two manifolds with complex metric that admit a second-order Killing tensor were completely classified by Koenigs in  1889 \cite{Koenigs}. These metrics were more carefully studied  and their algebraic properties quantified   by Kalnins et al. \cite{Winter,Kalnins}.  With the additional restriction that we are considering a real two metric  with conformal factor $V$  the functions in Table~\ref{fourCor} are included in this class.

\section{Comparison with type D Metrics}
\label{SECDDD}

All SAV metrics admitting second-order Killing tensors obtained in the previous analysis are of Petrov Type D. However the reverse statement is not true.
For completeness, we discuss the properties of the most general Type D metric. A complete list of Type D vacuum metrics was given by Kinnersley \cite{KinnPhDthesis}. He showed, to his surprise, that all type D vacuum metrics have at least two Killing vectors. In other words, they are either SAV spacetimes or admit two spacelike Killing vectors such as colliding plane waves. He postulated a profound connection between the existence of isometries such as these and the Petrov classification. A full understanding of the relationship of the Petrov classification and orbital structure of the spacetime has yet to be achieved.

The line element for the general type D spacetime can be written down and specific cases derived by various limiting procedures \cite{Plebanski197698,SolutionsToEinsteinsEQ,debever:1955}. The line element for the vacuum case is parametrized by four constants namely, the mass $m$ , the NUT parameter $l$ and 2 parameters, $\gamma$ and $\epsilon$ that are related to the angular momentum per unit mass:   
\begin{align}
ds^2 &=  \frac{p^2+q^2}{(1 - p q)^{2}} \left[ \frac{dq^2}{Y(q)} +\frac{dp^2}{X(p)}   \right]\notag\\
&+ \frac{X(p)(dt+q^2 d\phi)^2-Y(q) (  dt -p^2 d \phi)^2}{(1 - p q)^{2} (p^2+q^2) }, \label{metD}
\end{align}
where
\begin{align}
X(p) &= \gamma ( 1 -p^4) + 2 l p - \epsilon p^2 + 2m p^3, \notag\\
Y(q) &= \gamma ( 1 -q^4) - 2 m q + \epsilon q^2 - 2l q^3. 
\end{align}  
$X(p)$ must be positive to get a Lorentzian signature, and $Y(q)$ must be positive for axisymmetric stationary spacetimes.

We define the coordinates ($\hat{\rho}$, $\hat{z}$) such that $d \hat{\rho} = d q/\sqrt{Y(q)}$ and $d \hat{z} = d p / \sqrt{X(\rho)}$. Using these coordinates, the general type D metric can be cast in the form of Eq. \eqref{LineEle} with 
\begin{align} 
V_D  & =   \frac{p^2+q^2}{(1 - p q)^{2}},&R^2_D=\frac{X(p)Y(q)}{(p q -1)^4},\notag\\
 e^{2\psi} &=\frac{Y-X}{(p^2+q^2)(1 - p q)^{2}},   &\omega_D =\frac{p^2Y-q^2 X}{X-Y}. 
\end{align}
On the principal null tetrad, the only non vanishing component of the Weyl tensor is 
\begin{align}
\Psi_2 = -(m+i l)\left(\frac{1-pq}{q+ip}\right)^3.
\end{align}
Note that while the conformally rescaled metric $V_D g_D^{\alpha_1 \alpha_2} $ is separable, in general $ g_D^{\alpha_1 \alpha_2} $ is not. All Type D spacetimes are said to admit a conformal second-order Killing tensor, but as said previously,they fail to admit an ordinary second-order Killing Tensor.

The subset of type D spacetimes that admit a second-order Killing tensor can be obtained from Eq. \eqref{metD} by making the scale transformation \cite{SolutionsToEinsteinsEQ}
\begin{align}
p&\rightarrow n^{-1}p, & q&\rightarrow n^{-1}q, & \phi & \rightarrow n^3 \phi, & t\rightarrow n t, \notag\\
m+i l &\rightarrow n^{-3} (m+il), &\epsilon & \rightarrow n^{-2} \epsilon, & \gamma&\rightarrow n^{-4} \gamma.
\end{align}
and taking the limit of $n \rightarrow \infty$, yielding the metric derived in Eq. \eqref{SepMetFuncaa},
\begin{align}
ds^2 &=  (p^2+q^2) \left[ \frac{dq^2}{Y(q)} +\frac{dp^2}{X(p)}   \right]\notag\\
&+ \frac{X(p)(dt+q^2 d\phi)^2-Y(q) (  dt -p^2 d \phi)^2}{ (p^2+q^2) },
\end{align}
where
\begin{align}
X(p) &= \gamma    + 2 l p - \epsilon p^2, &
Y(q) &= \gamma   - 2 m q + \epsilon q^2.  \notag
\end{align}
The relationship between the variables $p$ and $q$, the constants $\gamma$, $l$, $m$ and $\epsilon$, and those used in  Table \ref {fourCor} are given in Eqs. \eqref{CONSTCON2} and \eqref{CONSTCON}.

\section{Conclusion}
\label{SECCON}
This paper provides a constructive method for calculating the second-order Killing Tensor components and identifying which SAV metrics admit these structures.

In conclusion, I will now comment more fully on the ways in which various steps in the calculation presented here will be generalized in subsequent work \cite{JdB3,JdBZV1}.  In a very real sense, the calculation performed here provides a prototype for the more complex calculations to come without distracting the reader with an excessive  proliferation of indices and other technical difficulties.

The restriction of the manifolds under consideration to SAV metrics  results in a line element \eqref{LineEle} that  can be split into two independent two-metrics. Namely, the components associated with the Killing vectors  indicated by the indices $(A,B)$ and the diagonal two-metric associated with the independent variables $(\rho,z)$ indicated by the indices $(i,j)$. This metric structure results in the decoupling of the Killing equations into two groups. As shown in Eq. \eqref{KT1}, the first group completely defines the gradients of the $T^{(AB)}$ components. The second group, Eqs. \eqref{KT2a}-\eqref{KT2d}, represent the Killing equations of a two manifold and are completely decoupled from the  $T^{(AB)}$ components. When searching for higher-order Killing tensors for SAV spacetimes, a similar decoupling into several groups takes place. The two groups identified in this example always persist. In other words, we always get a group of equations where the gradients of the Killing tensor components with indices totally in Killing vector directions, are fully defined, and a second group that contains the Killing equations for a two-manifold with metric $g_{ij}=V\delta_{ij}$. In addition, however, for higher-order Killing equations other groups of equations are also introduced allowing greater freedom in the solutions found as will be seen in \cite{JdB3}.

The analytic structure identified in Eqs. \eqref{CRS} and \eqref{KillT2sss} also persists to higher-order Killing tensor problems, since it is a feature of the Killing equations of a two-manifold \cite{Hall}.  For higher order problems, the gauge freedom in $R$ coupled with this analytic structure will also be exploited to simplify the equations and facilitate writing down explicit solutions.

The fact that the Killing equations can be cast in the symmetrical form given in Eqs.  \eqref{KillT2TABint} and \eqref{kill2Standard}, indicating that what is sought is not one but four different two-manifolds admitting a second-order Killing tensor \cite{JdB1}, came as a surprise.  It led the author to look for a similar structure for fourth order Killing tensors. Such a structure was found, greatly reducing the complexity of the problem \cite{JdB3}.

Eqs.  \eqref{KillT2TABint} and \eqref{kill2Standard} further imply that the metric functions are separable, and can be expressed in the forms \eqref{KILLT4VVVB} and \eqref{KILLT4VV}. It is this form of the functions that allows the Killing equations \eqref{KEEEQ}  to be solved formally, and an explicit closed-form solution \eqref{KILLT4s} to be written down, even though the exact forms of the metric functions $f$ have yet to be determined.
While the property of separability (of the metric functions and of the Hamilton-Jacobi equations) does not extend to higher-order Killing tensors in SAV spacetimes, it is possible to write down an explicit closed-form solution of the fourth order-Killing equations in terms of the metric functions. This reduction is carried out in \cite{JdBZV1}.

This paper highlights the origin of the four separable coordinate systems discovered by  Carter in SAV spacetimes that admit a second-order Killing tensor. They are classified in terms of  the analytic structure associated with the Killing equations. It is shown that the condition that a second-order Killing tensor exists on the spacetime is far more restrictive than the condition that the spacetime obeys the SAV field equations. In particular, Eqs. \eqref{KillT2TABint} and \eqref{kill2Standard} imply that the Killing equations impose a second-order linear differential equation on the field variables. The SAV field equations are much less restrictive; they are effectively fourth-order differential equations.  It is necessary to consider at least fourth-order Killing equations before the conditions for their existence place restrictions on the field variables  that are higher than fourth order. 
It can also be considered as somewhat artificial the circumstance that  a property of the spacetime should in some sense be dependent on a choice of gauge, as implied by the result that relates  the four separable coordinates to solutions of the Killing equations.  

There are several strong analytic indications as well as a considerable amount of numerical evidence \cite{JdB1} that higher-order and at least fourth order Killing tensors should be considered in order to obtain a full description of the orbital structure of SAV spacetimes. In subsequent papers \cite{JdB3,JdBZV1} higher-order Killing tensors will be more thoroughly explored.  The calculation increases considerably in complexity, however the basic approach is very similar to the example derivation of second-order Killing tensors presented in this paper. In fact many of the key ideas making the problem tractable were gleaned from this example.

\section{Acknowledgments}
My sincere thanks to Frank Estabrook for many useful discussions. I am also indebted to Tanja Hinderer and Michele Vallisneri for their insightful comments on the manuscript. I gratefully acknowledge support from NSF grants PHY-0653653, PHY-0601459, NASA grant NNX07AH06G, the Brinson Foundation and the David and Barbara Groce startup fund at Caltech. 

\appendix

\section{SAV Field Equations}
\label{FieldEQ}
For SAV spacetimes, the vacuum field equations for the quantities, $M_2$, $M_A$, $M_B$, and $M_C$, introduced in Sec. \ref{SEC2} are given by
\begin{align}
&M_{2,\zeta} = -M_2 M_2^*,   \notag\\
&M_{A,\zeta}=-M_AM_A^* -M_AM_2^* - M_2M_A^* -\frac{1}{4} M_BM_B^*\notag\\
&+  \frac{1}{2}\left(M_B( M_2^*+M_A^*)+M_B^*(M_2 +M_A)\right) -\frac{1}{4} M_CM_C^*,\notag\\
&M_{B,\zeta}= -\frac{1}{2}M_BM_B^* -2M_AM_A^*+ M_AM_B^*+M_BM_A^*\notag\\
& + (\frac{1}{2}M_B -M_A) M_2^*     + (\frac{1}{2}M_B^*-M_A^*)M_2 + \frac{1}{2}M_CM_C^*,\notag\\ 
& M_{C,\zeta}= -\frac{1}{2}M_CM_2^* -( \frac{3}{2}M_2 + 2M_A - M_B)M_C^*, \label{FieldEQMABC}
\end{align}
and 
\begin{align}
M_C^2&=4M_A^2 + 4M_2(M_A - M_B) - 4M_AM_B + M_B^2\notag\\&+M_{2 ,\overline{\zeta}}+2M_2^2.
\label{FieldEQMABCb}
\end{align}
This choice of variables was originally motivated as a linear combination of the variables introduced by Harrison \cite{HarrisonWET} and Neugebauer \cite{Neu} for use in the solution generation techniques. They are proportional to the rotation coefficients associated with the transverse frame of the SAV spacetime.

\section{Separable Coordinates }
\label{AppendSep}

This appendix specializes our formalism to the separable coordinate system, and gives the explicite form of the Killing equations, as well as the $M$ field variables expressed in terms of the separable functions. (The tilde over the $\tilde{M}$s, indicating the variables associated with the separable coordinate system is dropped here.)

In the separable coordinate system the Killing equations imply;
\begin{align}
  M_{2,\overline{\tilde{\zeta}}}  = CC, \notag\\
  2M_A^2+  M_{A,\overline{\tilde{\zeta}}}   =CC,\notag\\
 M_B^2 +M_{B, \overline{\tilde{\zeta}}}     = CC, \notag\\
(2M_A+M_B- M_2)M_C + M_{C, \overline{\tilde{\zeta}}}   = CC. \label{KillABCD}
\end{align}
The Killing equation for $M_2$ can be rewritten as,  
\begin{align}
\partial_{\tilde{\rho}\tilde{ z}} M_2=\partial_{\tilde{\rho}\tilde{ z}}M_2^* =0, \label{KillABCD2}
\end{align}
so the resulting $M_2$ must also have a separable form in this coordinate system or
\begin{align}
M_2 &= m_1(\tilde{\rho}) +im_2(\tilde{z}).
\end{align} 
This is consistent with the fact that for all coordinate systems considered in Table \ref{fourCor} the gauge function $R$ can be expressed as the product $R=r_1(\tilde{\rho}) r_2(\tilde{z})$.

Note that throughout this paper we follow the convention that functions of $\tilde{\rho}$ only, are indicated by a function subscript of 1, for example   $f_1(\tilde{\rho})$,   functions of $\tilde{z}$ only
by a  function subscript of 2.

The field variables  $M$ can also be expressed in terms of the separable functions as 
\begin{align}
M_A&=\frac{1}{2}\frac{f_1^{\phi\phi'}+if_2^{\phi\phi'}}{f_1^{\phi\phi}+f_2^{\phi\phi}},\notag\\
M_B&= \frac{f_1'+if_2'}{f_1+f_2},\notag\\
M_C 
 &=\left(  \frac{f_1^{\phi\phi'}+if_2^{\phi \phi'}}{  f_1^{\phi\phi}+f_2^{\phi \phi}  } -  \frac{f_1^{t\phi'}+if_2^{t \phi'}}{  f_1^{t\phi}+f_2^{t \phi}  }  \right) \frac{f_1^{t\phi}+f_2^{t \phi}}{f_1+f_2  } R .  \label{SepMetFunc}
\end{align}

\section{Computation of the functional form of metric functions in separable coordinates.}
\label{SeperableFunctions}
This appendix details the computation of the functional form of the separable metric functions that obey both the second-order Killing and the field equations. While this calculation  in principle appears straightforward,  given the equations and separable functions of the previous two appendices, it turned out to be unexpectedly tedious to execute.  In this section we detail the crucial simplifying steps and arguments that facilitate obtaining the functional form from first principles.  For the purpose of  illustrating useful concepts and the generalization to higher-order Killing tensors, this appendix is of very little importance. It is given mainly to prevent the reader from musing about useless details by providing them explicitly. 

The second and third equations of  \eqref{FieldEQMABC} can be combined to completely decouple $M_C$. Denoting $M_D = M_B-2M_A$, the resulting differential equation is
\begin{align}
2 M_{B,\zeta}- M_{D,\zeta} +M_DM_D^* =  \frac{3}{2}(M_DM_2^* +M_D^*M_2). \label{eqMDB}
\end{align}
The third equation  in  \eqref{FieldEQMABC} that defines $M_{B,\zeta}$ can also be  expressed in terms of $M_D$ as follows 
\begin{align}
2 M_{B,\zeta} +M_DM_D^* = M_DM_2^* +M_D^*M_2 + M_CM_C^*. \label{eqMBzMD}
\end{align}
Differentiating Eq. \eqref{eqMBzMD}
with respect to $\zeta$ and substituting both the Killing and Field equations leads to 
\begin{align}
M_{B,\zeta\zeta}+M_{B,\zeta}( 2M_B^*-\frac{3}{2}M_2^*)
+\frac{1}{2}M_2 M_{B,\overline{\zeta}}
\notag\\ 
-\frac{3}{2}M_BM_{2,\overline{\zeta}} +M_2(-3M_2M_B+2M_B^2-\frac{3}{2}M_2^*M_B^*)\notag\\
+ (2M_2)M_{2,\overline{\zeta}} +M_2(M_2^2+M_2^{*2})=0. \label{EQMBKill2}
\end{align}
Substituting  the separable function expressions for $M_B$ and $M_2$ into Eq. \eqref{EQMBKill2} for the four possible transformations, and repeatedly differentiating the resulting equations, the following conditions  are obtained for the separable functions:

\begin{align}
A),B)&&\frac{d^5 f_2}{d\tilde{z}^5}=\frac{d^5 f_1}{d\tilde{\rho}^5}=0, \\
C),D)&&\frac{d^3 f_1}{d x ^3}=\frac{d^3 f_2}{d y ^3} =0
\end{align}
The  transformation to $(x,y)$ coordinates for the cases $C$ and $D$ are given in Table \ref{fourCor}.
These equations, along with the consistency conditions that arise when the functions are substituted back into Eq. \eqref{EQMBKill2}, imply the functions listed in Table   \ref{fourCor}.

In addition, a number of integrability conditions  arise when repeatedly differentiating the field Eqs. \eqref{FieldEQMABC} \eqref{FieldEQMABCb} and Killing Eqs \eqref{KillABCD}. These equations  limit the freedom of the constants that appear in Table   \ref{fourCor}. In particular, in order for field and Killing equations to be consistent, they must satisfy the condition  
\begin{align}
g^2+ b\ g + c = 0, \label{ConsistencyCond}
\end{align}
where 
\begin{align}
g &= \frac{1}{4V}(M_2 M_2^*-2M_{B , \overline{\zeta}}^*),\notag\\
b&= -\frac{1}{2V}(M_2-M_B)(M_2^*-M_B^*),\notag\\
c&=\frac{1}{4}(b^2- e \overline{e}) ,\notag\\
e&= \frac{1}{2V}(M_2^2+M_2^{* 2} + M_B^{* 2} +2M_{2, \overline{\zeta}})\notag\\
&- \frac{1}{V}(   M_2 M_B + M_2^* M_B^*).
\end{align}
Eqs. \eqref{ConsistencyCond} result in the constraints on the constants listed in Table   \ref{fourCor}.

Once the functions $f_1$ and $f_2$ have been determined, the general form of the separable functions corresponding to the $t, \ \phi$ components of the metric can be written down using the following argument.
The determinant of the metric $\mbox{det}(g_{AB})=-R^2$ 
relates the remaining separable metric functions to the 
functions  $f_1$, $f_2$ and $R$, already found with;  
\begin{align}
 \frac{(f_1+f_2)^2}{R^2}=-( f_1^{\phi\phi}+f_2^{\phi\phi} ) ( f_1^{tt}+f_2^{tt}) +(f_1^{t\phi}+f_2^{t\phi})^2.  
\end{align} 
By differentiating with respect to $\tilde{\rho}$ and $\tilde{z}$, and recalling that a function with subscript $1$ is a function of $\tilde{\rho}$ only whereas a subscript $2$ indicates a function of $\tilde{z}$, we have that
\begin{align}
\partial_{\tilde{\rho}\tilde{z}} F(\tilde{\rho},\tilde{z}) = - f_1^{\phi\phi'} f_2^{tt'} -f_2^{\phi\phi'} f_1^{tt'} +2f_1^{t\phi'}f_2^{t\phi'}, \label{EWD}  
\end{align}
where $F(\tilde{\rho},\tilde{z})=  (f_1+f_2)^2/R^2 $. To find the functional form of $f_1^{ \phi\phi'}$, for example, divide \eqref{EWD} by the coefficient of  $f_1^{ t \phi'}$ and differentiate w.r.t. $\tilde{z}$, removing any dependence on   $f_1^{ t \phi'}$. Divide the resulting equation by the coefficient of   $f_1^{ t t'}$,  differentiate w.r.t. $\tilde{z}$ and solve for $f_1^{ \phi\phi'}$. This process can be repeated for any of the six metric functions, and indicates that the correct functional form for the functions are
\begin{align}
f^{AB}_1(\tilde{\rho})= \left. D_0^{AB} +\Sigma_{j=1}^3 D_j^{AB} \partial_{\tilde{z}}^j  F(\tilde{\rho},\tilde{z}) \right|_{\tilde{z} = z_0},  \notag\\
f^{AB}_2(\tilde{z})= \left. E_0^{AB} +\Sigma_{j=1}^3E_j^{AB} \partial_{\tilde{\rho}}^j  F(\tilde{\rho},\tilde{z}) \right|_{\tilde{\rho} = \rho_0}.
\end{align}
The remaining metric functions are thus determined up to a set of unknown constants  $D_j^{AB}$ and $E_j^{AB}$ . These functions are then substituted back into the field equations to determine the constants in terms of the constants that enter the known functions, $f_1$, $f_2$ and $R$.

The results can be most concisely expressed using 
\begin{align} 
V_D  & =  p^2+q^2=f_1+f_2, &R^2_D&=X(p)Y(q).
\end{align}
The definitions of $p$ and $q$ for the different analytic structures (A.S.) are given by
\begin{align}
\begin{array}{|c|c|c|}\hline\hline
\mbox{A. S.}&  q & p   \\ \hline
(A)&    \sqrt{b_4}\left(  \tilde{\rho}^2 - \frac{b_2}{2b_4} \right)  &  \sqrt{2b_2} \left(\tilde{z} +\frac{b_1}{4 b_2}\right)    \\ \hline
(B)&     \sqrt{b_4}\left(\tilde{\rho}^2 + \frac{b_2}{2b_4}\right)&  \sqrt{c_4} \left(\tilde{z}^2 + \frac{ b_2}{2c_4}  \right)       \\ \hline
(C)&     \sqrt{ b_4}\left( x + \frac{ b_2}{ 2 b_4  }  \right) &  \sqrt{2c_2}\left( y+\frac{c_4}{4c_2} \right)   \\ \hline
(D)&   \sqrt{ b_2}\left(x+ \frac{b_1}{2b_2} \right) & \sqrt{c_2}\left(y+\frac{c_1}{2c_2}\right)   \\ \hline
\end{array}
 \label{CONSTCON2}
\end{align}
Functions $X$ and $Y$ are defined   for each of the four coordinate systems  given in Table \ref{fourCor} as
\begin{align} 
X(p) &= \gamma    + 2 l p - \epsilon p^2,& 
Y(q) &= \gamma   - 2 m q + \epsilon q^2; 
\end{align}
the relationship between the constants $\gamma$, $l$, $m$ and $\epsilon$ and the constants entering Table \ref{fourCor} for the four different analytic structures are
\begin{align}
\begin{array}{|c|ccll|}\hline\hline
(A)&\epsilon = 0 &\gamma = 2   b_2&m = -2\sqrt{b_4} &l=0     \\ \hline
(B)&\epsilon=0 &\gamma=-b_2 &m =-2 \sqrt{b_4}     &l = 2\sqrt{c_4}      \\ \hline
(C)&\epsilon = \frac{1}{4}&\gamma = \frac{1}{4}\frac{b_2^2}{4b_4}=\frac{c_2}{2}-\frac{c_4^2}{4\cdot 8 c_2  }   &m=\frac{1}{8} \frac{b_2}{\sqrt{b_4}}&l= \frac{c_4}{2 \sqrt{2c_2}  }   \\ \hline
(D)&\epsilon = 1&\gamma = \frac{b_1^2-4b_2^2}{4b_2}=\frac{4c_2^2-c_1^2}{4c_2}&m = \frac{b_1}{2\sqrt{b_2}} &l = \frac{c_1}{2\sqrt{c_2}}  \\ \hline 
\end{array}\label{CONSTCON}
\end{align}

The separable functions entering the metric thus are 
\begin{align}
f^{\phi\phi}_1 &= -\frac{1}{Y}, & f^{\phi\phi}_2 &= \frac{1}{X}, &
f^{t\phi}_1 &= \frac{q^2}{Y}, \notag\\
f^{tt}_1 &= -\frac{q^4}{Y}, & f^{tt}_2 &= \frac{p^4}{X},& f^{t\phi}_2 &= \frac{p^2}{X}, \label{SepMetFuncX}
\end{align}
and the resulting metric becomes
\begin{align}
ds^2 &=  (p^2+q^2) \left[ \frac{dq^2}{Y(q)} +\frac{dp^2}{X(p)}   \right]\notag\\
&+ \frac{X(p)(dt+q^2 d\phi)^2-Y(q) (  dt -p^2 d \phi)^2}{ (p^2+q^2) }. \label{SepMetFuncaa}
\end{align}

\bibliographystyle{apsrev}

\bibliography{../BholesNemadon}

\begin{thebibliography}{23}
\expandafter\ifx\csname natexlab\endcsname\relax\def\natexlab#1{#1}\fi
\expandafter\ifx\csname bibnamefont\endcsname\relax
  \def\bibnamefont#1{#1}\fi
\expandafter\ifx\csname bibfnamefont\endcsname\relax
  \def\bibfnamefont#1{#1}\fi
\expandafter\ifx\csname citenamefont\endcsname\relax
  \def\citenamefont#1{#1}\fi
\expandafter\ifx\csname url\endcsname\relax
  \def\url#1{\texttt{#1}}\fi
\expandafter\ifx\csname urlprefix\endcsname\relax\def\urlprefix{URL }\fi
\providecommand{\bibinfo}[2]{#2}
\providecommand{\eprint}[2][]{\url{#2}}

\bibitem[{\citenamefont{Carter}(1968)}]{CarterSeparability}
\bibinfo{author}{\bibfnamefont{B.}~\bibnamefont{Carter}},
  \bibinfo{journal}{Commun.\ Math.\ Phys.} \textbf{\bibinfo{volume}{10}},
  \bibinfo{pages}{280} (\bibinfo{year}{1968}).

\bibitem[{\citenamefont{Walker and Penrose}(1970)}]{Walker}
\bibinfo{author}{\bibfnamefont{M.}~\bibnamefont{Walker}} \bibnamefont{and}
  \bibinfo{author}{\bibfnamefont{R.}~\bibnamefont{Penrose}},
  \bibinfo{journal}{Commun. Math. Phys} \textbf{\bibinfo{volume}{18}},
  \bibinfo{pages}{265} (\bibinfo{year}{1970}).

\bibitem[{\citenamefont{Chandrasekhar}(1983)}]{MathTheoryofBlackHoles}
\bibinfo{author}{\bibfnamefont{S.}~\bibnamefont{Chandrasekhar}},
  \emph{\bibinfo{title}{The Mathematical Theory of Black Holes}}
  (\bibinfo{publisher}{Clarendon Press. Oxford}, \bibinfo{year}{1983}).

\bibitem[{\citenamefont{Brink}(2008{\natexlab{a}})}]{JdB0}
\bibinfo{author}{\bibfnamefont{J.}~\bibnamefont{Brink}},
  \bibinfo{journal}{Phys.\ Rev.\ D} \textbf{\bibinfo{volume}{78}},
  \bibinfo{eid}{102001} (\bibinfo{year}{2008}{\natexlab{a}}).

\bibitem[{\citenamefont{Drasco and Hughes}(2006)}]{WaveFormMachineSteveDrasco}
\bibinfo{author}{\bibfnamefont{S.}~\bibnamefont{Drasco}} \bibnamefont{and}
  \bibinfo{author}{\bibfnamefont{S.~A.} \bibnamefont{Hughes}},
  \bibinfo{journal}{Phys.\ Rev.\ D} \textbf{\bibinfo{volume}{D 73}},
  \bibinfo{pages}{024027} (\bibinfo{year}{2006}).

\bibitem[{\citenamefont{Brink}(2008{\natexlab{b}})}]{JdB1}
\bibinfo{author}{\bibfnamefont{J.}~\bibnamefont{Brink}},
  \bibinfo{journal}{Phys.\ Rev.\ D} \textbf{\bibinfo{volume}{78}},
  \bibinfo{eid}{102002} (\bibinfo{year}{2008}{\natexlab{b}}).

\bibitem[{\citenamefont{Brink}(2009{\natexlab{a}})}]{JdB3}
\bibinfo{author}{\bibfnamefont{J.}~\bibnamefont{Brink}}, \bibinfo{journal}{IV
  (in preparation), Relationship between Weyl Curvature and Killing Tensors in
  SAV Spacetimes}  (\bibinfo{year}{2009}{\natexlab{a}}).

\bibitem[{\citenamefont{Brink}(2009{\natexlab{b}})}]{JdBZV1}
\bibinfo{author}{\bibfnamefont{J.}~\bibnamefont{Brink}},
  \bibinfo{journal}{Formal solution of the Fourth Order Killing tensors
  equations of SAV Spacetimes.}  (\bibinfo{year}{2009}{\natexlab{b}}).

\bibitem[{\citenamefont{Brink}(2009{\natexlab{c}})}]{JdBZV2}
\bibinfo{author}{\bibfnamefont{J.}~\bibnamefont{Brink}},
  \bibinfo{journal}{Poincare Maps of Static Spacetimes with Equatorial
  Symmetry-Example Zipoy Voorhees Metric.}
  (\bibinfo{year}{2009}{\natexlab{c}}).

\bibitem[{\citenamefont{Koenigs}(1972)}]{Koenigs}
\bibinfo{author}{\bibfnamefont{G.}~\bibnamefont{Koenigs}},
  \bibinfo{journal}{Lecons sur la theorie generale des surfaces.}
  \textbf{\bibinfo{volume}{4}}, \bibinfo{pages}{368} (\bibinfo{year}{1972}).

\bibitem[{\citenamefont{Whittaker}(1944)}]{Whit}
\bibinfo{author}{\bibfnamefont{E.}~\bibnamefont{Whittaker}},
  \emph{\bibinfo{title}{A Treatize on the Analytical Dynamics of particles and
  rigid bodies}} (\bibinfo{publisher}{New York Dover Publications},
  \bibinfo{year}{1944}).

\bibitem[{\citenamefont{Wolf}(1998)}]{WolfT}
\bibinfo{author}{\bibfnamefont{T.}~\bibnamefont{Wolf}}, \bibinfo{journal}{Comp.
  Phys. Comm} pp. \bibinfo{pages}{316--329} (\bibinfo{year}{1998}).

\bibitem[{\citenamefont{Hall}(1983)}]{Hall}
\bibinfo{author}{\bibfnamefont{L.~S.} \bibnamefont{Hall}},
  \bibinfo{journal}{Physica D} \textbf{\bibinfo{volume}{8}},
  \bibinfo{pages}{90} (\bibinfo{year}{1983}).

\bibitem[{\citenamefont{Hietarinta}(1987)}]{Hietarinta}
\bibinfo{author}{\bibfnamefont{J.}~\bibnamefont{Hietarinta}},
  \bibinfo{journal}{Physics Reports} \textbf{\bibinfo{volume}{147}},
  \bibinfo{pages}{87} (\bibinfo{year}{1987}).

\bibitem[{\citenamefont{Kalnins et~al.}(2005)\citenamefont{Kalnins, Kress, and
  Miller}}]{Kalnins}
\bibinfo{author}{\bibfnamefont{E.~G.} \bibnamefont{Kalnins}},
  \bibinfo{author}{\bibfnamefont{J.~M.} \bibnamefont{Kress}}, \bibnamefont{and}
  \bibinfo{author}{\bibfnamefont{W.}~\bibnamefont{Miller}},
  \bibinfo{journal}{J. Math. Phys.} \textbf{\bibinfo{volume}{46}},
  \bibinfo{pages}{053509} (\bibinfo{year}{2005}).

\bibitem[{\citenamefont{Xanthopoulos}(1984)}]{Xanth3}
\bibinfo{author}{\bibfnamefont{B.~C.} \bibnamefont{Xanthopoulos}},
  \bibinfo{journal}{J. Phys. A: Math. Gen} \textbf{\bibinfo{volume}{17}},
  \bibinfo{pages}{87} (\bibinfo{year}{1984}).

\bibitem[{\citenamefont{Kalnins et~al.}(2002)\citenamefont{Kalnins, Kress, and
  Winternitz}}]{Winter}
\bibinfo{author}{\bibfnamefont{E.~G.} \bibnamefont{Kalnins}},
  \bibinfo{author}{\bibfnamefont{J.~M.} \bibnamefont{Kress}}, \bibnamefont{and}
  \bibinfo{author}{\bibfnamefont{P.}~\bibnamefont{Winternitz}},
  \bibinfo{journal}{J. Math. Phys.} \textbf{\bibinfo{volume}{43}},
  \bibinfo{pages}{970} (\bibinfo{year}{2002}).

\bibitem[{\citenamefont{Kinnersley}(1968)}]{KinnPhDthesis}
\bibinfo{author}{\bibfnamefont{W.~M.} \bibnamefont{Kinnersley}},
  \bibinfo{journal}{Ph.D. thesis, CALTECH}  (\bibinfo{year}{1968}).

\bibitem[{\citenamefont{Plebanski and Demianski}(1976)}]{Plebanski197698}
\bibinfo{author}{\bibfnamefont{J.~F.} \bibnamefont{Plebanski}}
  \bibnamefont{and}
  \bibinfo{author}{\bibfnamefont{M.}~\bibnamefont{Demianski}},
  \bibinfo{journal}{Annals of Physics} \textbf{\bibinfo{volume}{98}},
  \bibinfo{pages}{98 } (\bibinfo{year}{1976}).

\bibitem[{\citenamefont{Hoenselaers and Dietz}(1983)}]{SolutionsToEinsteinsEQ}
\bibinfo{editor}{\bibfnamefont{C.}~\bibnamefont{Hoenselaers}} \bibnamefont{and}
  \bibinfo{editor}{\bibfnamefont{W.}~\bibnamefont{Dietz}}, eds.,
  \emph{\bibinfo{title}{Solutions of Einstein's Equations: Techniques and
  Results, Proceedings of the International Seminar on Exact Solutions of
  Einsteins Equations}} (\bibinfo{publisher}{Springer-Verlag},
  \bibinfo{year}{1983}).

\bibitem[{\citenamefont{Debever et~al.}(1984)\citenamefont{Debever, Kamran, and
  McLenaghan}}]{debever:1955}
\bibinfo{author}{\bibfnamefont{R.}~\bibnamefont{Debever}},
  \bibinfo{author}{\bibfnamefont{N.}~\bibnamefont{Kamran}}, \bibnamefont{and}
  \bibinfo{author}{\bibfnamefont{R.~G.} \bibnamefont{McLenaghan}},
  \bibinfo{journal}{Journal of Mathematical Physics}
  \textbf{\bibinfo{volume}{25}}, \bibinfo{pages}{1955} (\bibinfo{year}{1984}),
  \urlprefix\url{http://link.aip.org/link/?JMP/25/1955/1}.

\bibitem[{\citenamefont{Harrison}(1983)}]{HarrisonWET}
\bibinfo{author}{\bibfnamefont{B.~K.} \bibnamefont{Harrison}},
  \bibinfo{journal}{J. Math. Phys.} \textbf{\bibinfo{volume}{24}},
  \bibinfo{pages}{2178} (\bibinfo{year}{1983}).

\bibitem[{\citenamefont{Neugebauer}(1979)}]{Neu}
\bibinfo{author}{\bibfnamefont{G.}~\bibnamefont{Neugebauer}},
  \bibinfo{journal}{J. Phys. A: Math. Gen} \textbf{\bibinfo{volume}{12}},
  \bibinfo{pages}{L67} (\bibinfo{year}{1979}).

\end{thebibliography}

\end{document}